\documentstyle[aps,prd,twocolumn,epsfig]{revtex}
\def\b{\bar}
\def\d{\partial}

\def\cD{{\cal D}}

\def\m{\mu}
\def\n{\nu}

\def\t{\tau}

\def\~{\tilde}

\def\bY3{\bar Y_{,3}}
\def\Y3{Y_{,3}}
\def\z{\zeta}
\def\Z{{\b\zeta}}
\def\Y{{\bar Y}}
\def\cZ{{\bar Z}}
\def\`{\dot}
\def\be{\begin{equation}}
\def\ee{\end{equation}}
\def\bea{\begin{eqnarray}}
\def\eea{\end{eqnarray}}

\def\fn{\footnote}

\def\cF{{\cal F}}

\def\mn{{\mu\nu}}

\begin{document}
\title{Electromagnetic Excitation of Rotating Black Holes and
Relativistic Jets}

\author{Alexander Burinskii\\
Gravity Research Group, NSI, Russian Academy of
Sciences, \\
B. Tulskaya 52  Moscow 115191 Russia;}

\author{Emilio Elizalde\\
ICE Consejo Superior de Investigaciones Cient\'{\i}ficas and IEEC,\\ \&
Campus UAB, Facultat de Ci\`encies \\ Torre C5-Parell-2a planta,
E-08193 Bellaterra (Barcelona), Spain;}

\author{Sergi R. Hildebrandt\\
Instituto de Astrofisica de Canarias,\\
C/Via Lactea s/n, La Laguna, Tenerife, 38200, Spain;}

\author{Giulio Magli \\
Dipartimento di Matematica del Politecnico di Milano,\\
Piazza Leonardo Da Vinci 32, 20133 Milano, Italy.}
\maketitle
\begin{abstract}
We show that electromagnetic excitations of rotating black holes
can lead to the appearance of narrow singular beams which break up
the black hole horizon forming a tube-like region which connects
the interior and exterior. It is argued that this effect may be at
the origin of jet formation.
\end{abstract}

\medskip

{\bf 1. Introduction.}

\medskip

The Kerr solution describes the stationary phases of rotating
blackholes. Non-stationary behaviors, like bursts and jets, should
of course be described by nonstationary solutions which, however,
are not known so far. On the other hand, the Kerr metric is only
one representative of the broad Kerr-Schild class of
Einstein-Maxwell solutions. Among these solutions there are
rotating Kerr-Schild solutions containing axial, semi-infinite
singular lines (beams). These solutions have not been paid so far
enough attention in astrophysical applications, and thus have
never been analyzed in detail from the physical point of view.
These solutions however may be considered as a low frequency limit
of some wave solutions related to the aligned (coherent)
excitations of the rotating black holes. In the paper \cite{BEHM1}
the structure of horizons for the stationary solutions with axial
singularities was considered  and  some of the astrophysical
effects which may be related to the appearance of the axial
singularities. It was shown that physical consequences originated
from these singularities turn out to be crucial for the initially
stable black holes. Even weak axial singularities ``break up" the
Kerr black hole, forming a "hole in the horizon". As a result, the
internal region turn out to be connected with the external one,
and the black hole turns out to be ``half dressed''.

In this paper we  obtain the exact wave solutions of the Maxwell
equations on the Kerr background which are aligned to the Kerr
principal null congruence and show that these solutions are the
wave generalizations of the corresponding exact stationary
Kerr-Schild solutions. In the quasi-stationary low frequency
limit, they tend to the exact self-consistent solutions with an
arbitrary degree of approximation in the whole space-time but for
the excision of a narrow vicinity of the beam. This leads us to
the conclusion that the appearance of the holes in horizon and
beams may be caused by a coherent excitation of the rotating
sources \cite{BEHM1}. The corresponding physical processes  may
result in the production of jets \cite{jets}.

\medskip

{\bf 2. Solutions with axial singularities.}

\medskip

In the fundamental paper by Debney, Kerr and Schild \cite{DKS}
exact solutions were obtained for the metric form \be g^{\mn}=\eta
^{\mn} - 2H k^\m k^\n , \label{ksa}\ee where $k^\m$ is a
(\ref{ksa}) with a null vector field $k^\m$ ($k_\m k^\m =0$),
which is tangent to a geodesic and shear-free principal null
congruence (PNC). These spacetimes are algebraically special; as a
consequence, many tetrad Ricci components vanish and there is a
strong restriction on the tetrad components of the electromagnetic
field. The principal property of these solutions is that the
electromagnetic field is aligned with the Kerr PNC,  satisfying
the constraints \be F^\mn k_\m=0 \label{align}.\ee It  is
described by two non-vanishing tetrad components of the self-dual
tensor $\cF_{12} =AZ^2 , \quad \cF _{31} =\gamma Z - (AZ),_1 , $
where commas denote the directional derivatives wrt the chosen
null tetrad vectors.\fn{The real null tetrad vector $e^3 \equiv
e_4 = k_\m dx^\m $.} The resulting equations for the e.m.  field
are \be A,_2-2 Z^{-1} \cZ Y,_3 A = 0, \quad \gamma ,_4=0
,\label{3}\ee \be \cD A+ \cZ ^{-1} \gamma ,_2 - Z^{-1} Y,_3 \gamma
=0 . \label{4}\ee

Gravitational field equations yield \be M,_2 - 3 Z^{-1} \cZ Y,_3 M
= A\bar\gamma \cZ , \quad \cD M  = \frac 12 \gamma\bar\gamma  ,
\label{6}\ee where $ \cD=\d _3 - Z^{-1} Y,_3 \d_1 - \cZ ^{-1} \Y
,_3 \d_2   \ . $

Solutions of this system were given in \cite{DKS} only for the
stationary case, with $\gamma=0$. We show that the equations for
e.m. field may be integrated for  $\gamma \ne 0$. To get an
oscillating solution, one defines a complex retarded-time
parameter $\t = t -r +ia\cos \theta$ which
  satisfies the relations $ (\t),_2=(\t),_4=0.$
  It allows one to represent (\ref{3}) in the form \cite{Bur-nst}
\be (AP^2),_2=0, \quad P,_2=-P Y,_3 \label{8} .\ee
 This equation can be integrated,
yielding $ A=\psi(Y,\t)/P^2 $. It has the form obtained in
\cite{DKS}. The only difference is in the extra dependence of the
function $\psi$ from the retarded-time parameter $\t$. It means,
that the stationary solutions obtained in \cite{DKS} may be
considered as low-frequency (or adiabatical) limits of these
solutions.

One can check that the action of the operator $\cD$ on the
variables $Y, \bar Y $ and $ \rho =x^\m e^3_\m$ is  \be \cD Y =
\cD \bar Y = 0,\qquad \cD \rho =1 \ , \label{10}\ee and therefore
$\cD \rho = \d \rho / \d t_0 \cD t_0  = P\cD t_0 =1 $, what yields
\be \cD t_0 = P^{-1} . \ee As a result, Eqs.~(\ref{4}) take the
form \be \dot A = -(\gamma P),_{\bar Y} ,\quad \gamma,_4=0 ,
\label{11} \ee where $\dot {( \ )} \equiv \d_{t_0}$.

For the stationary background  considered here,
$P=2^{-1/2}(1+Y\bar Y)$, and $\dot P = 0$.  The coordinates $Y$,
and $\t$ are independent from $\bar Y$, which allows us to
integrate (\ref{11}). We obtain the following general solution \be
\gamma = - P^{-1}\int \dot A d\bar Y =
 \frac{2^{1/2}\dot \psi} {P^2 Y} +\phi (Y,\t)/P ,
\label{12}\ee where $\phi$ is an arbitrary analytic function of
$Y$ and $\t$. The term $\gamma$  in $ \cF _{31} =\gamma Z -
(AZ),_1  \ $ describes a part of the null electromagnetic
radiation which falls off asymptotically as $1/r$ and propagates
along the Kerr principal null congruence $e^3$. It follows from
(\ref{6}) that $\gamma$ describes a loss of mass by radiation with
the stress-energy tensor $\kappa T^{(\gamma)}_\mn = \frac 12
\gamma \bar \gamma e^3_{\m} e^3_{\n}$.

We now evaluate the term $(AZ),_1. $
 For the stationary case we have the relations $Z,_1
=2ia \bar Y (Z/P)^3 $ and  $\t,_1 =- 2ia \bar Y Z/P^2 $ . This
yields \be (AZ),_1 = \frac{Z}{P^2} (\psi ,_Y - 2ia  \dot \psi
\frac{\bar Y}{P^2} - 2 \psi \frac{P_Y}{P}) + A 2ia \frac{Z \bar
Y}{P^3} . \label{AZ1} \ee Since $Z/P =1/(r+ia \cos \theta)$, this
expression contains terms which fall off like $r^{-2}$ and
$r^{-3}$. However, it contains also factors which depend on the
coordinate $Y = e^{i\phi} \tan \frac {\theta} 2 $ and can be
singular at the $z$-axis, forming narrow beams, i.e. the
half-infinite lines of singularity. In particular, these lines can
be the $z^+$ or $z^-$ axis, which correspond to $\theta =0$ and
$\theta=\pi$ (cases $n=\pm1$, respectively).

The exact Debney-Kerr-Schild solutions arise in the limit $\gamma=
0 ,$ which yields a constant electromagnetic field.  The unique
non-zero component of the field tensor in this case is $\cF _{31}
= -(AZ),_1$ where $Z$ is the (complex) expansion of the PNC.  The
function $A$ has the general form \be A= \psi(Y)/P^2, \ee where
$P=2^{-1/2}(1+Y\Y)$, and $\psi$ is an arbitrary holomorphic
function of $Y$.  The resulting metric has the Kerr-Schild form
(\ref{ksa}), where the function $h$ is given by \cite{DKS} \be
h=m(Z+\bar Z)/(2P^3) - A \bar A Z\bar Z /2. \ee In terms of
spherical coordinates on the flat background one has $Y(x) =
e^{i\phi} \tan \frac {\theta} 2$, which is singular at
$\theta=\pi$. This singularity will be present in any holomorphic
function $\psi (Y)$, and, consequently, in $A$ and in $h$.
Therefore, all the solutions of this class
---with the exclusion of the case $\psi =$ const which corresponds
to the Kerr-Newman solution--- will be singular at some angular
direction $\theta$.

The simplest cases are $\psi=q/(Y +c)$ and $\psi=q(Y +b)/(Y+c)$,
which correspond to an arbitrary direction of the axial
singularity. However, the sum of singularities in different
directions is also admissible $\psi(y)=\sum_i q_i(Y
+b_i)/(Y+c_i)$, as well as polynomials of higher degree.
 Notice, that the axial singularity  survives for arbitrarily
small values of $q$ when $q\ne0.$ It yields the solutions which tend
to the exact one in the limit $q/m\to0$ everywhere but for the
exclusion of an $\epsilon$-vicinity of the axial singularity.

In the quasi-stationary limit, $\dot A \to 0,$ from Eq.(\ref{11})
it follows that the solutions correspond to the well known
$\gamma=0$ solutions.

\medskip

{\bf 3. Causal structure.}

\medskip

The properties of the horizons of these solutions were considered
in \cite{BEHM1} and it was shown that the black holes turn out to
be broken up by the axial singularities, with the appearance of a
tubelike region which connects internal and external regions,
allowing matter to escape. As a result, the circular Kerr
singularity turns out to be ``half-dressed''. These solutions do
not contradict the assertion of the ``No hair theorem" of Carter
and Robinson, which states uniqueness of the Kerr and Kerr-Newman
solutions \cite{Cha,MTW} under certain {\it regularity}
hypotheses, since these hypothesis are here {\it not} satisfied
\cite{BEHM1}. The structure of the horizons  for the solutions
containing axial singular lines follows from the form of metric
(\ref{ksa})
 where $k^\m$ is a null vector field $k_\m k^\m =0$
which is tangent to the Kerr principal null congruence. The
function $H$ has the form \be H= \frac {mr-e^2/2} {r^2 + a^2
\cos^2 \theta},\ee where the oblate coordinates $r,\theta$ are
used on the flat Minkowski background $\eta^{\mn}$. In the case of
rotating Kerr solutions, the Schwarzschild horizon splits into
four surfaces. Two surfaces correspond to the
staticity limit, $ r_{s+}$ and
$r_{s-}$, which are determined by the condition $g_{00} =0$, and
two surfaces come from the event and Cauchy horizons,
$S(x^\m)=$ const.,
which are the null surfaces determined by the condition $ g^\mn
(\partial _\m S) (\partial _\m S) =0. $ In the case $e^2 +a^2
>M^2$, the horizons of the Kerr-Newman solution disappear and the
Kerr singular ring turns out to be naked. The simplest axial
singularity is the pole $\psi =q/Y.$ In this case the boundaries
of the ergosphere, $r_{s+}$ and $r_{s-}$, are determined by the
condition $g_{00} =0$  and the solution acquires a new feature:
the surfaces $r_{s+}$ and $r_{s-}$ turn out to be joined by a
tube, forming a simply connected surface.

The surfaces of the {\it event horizons} are null and obey the
differential equation  \be (\d_r S)^2 [ r^2 +a^2 +(q/\tan \frac
\theta 2)^2 -2Mr ] - (\d_{\theta} S)^2 =0 \label{horh}. \ee
\begin{figure}[ht]
\centerline{\epsfig{figure=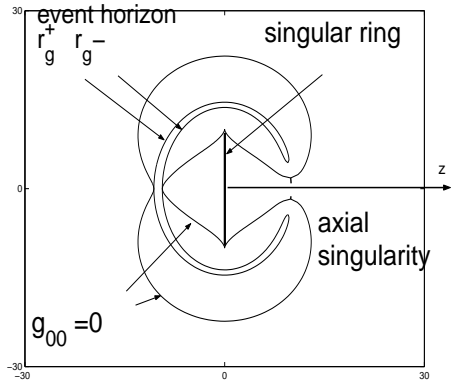,height=5.1cm,width=6cm}}
\caption{{\protect\small  Near extremal black hole with a hole in the
horizon, for
$m=10, \  a=9.98, \ q=0.1$. The event horizon is a closed connected
surface surrounded by the closed connected surface $g_{00}=0$. }}
\end{figure}
Similar to the boundary of the ergosphere, the two event horizons
are joined into one connected surface, and the surface of the
event horizon lies inside the boundary of the ergosphere. The
resulting surfaces are shown in Figs.~1, while other examples can
be found in \cite{BEHM1}. As it is seen from the figure, the axial
singularities lead to the formation of the holes in the black hole
horizon, which opens the interior of the ``black hole'' up to
external space.

The structure of the diagrams of the maximal analytic extension
(MAE) was also discussed in \cite{BEHM1}. It depends on the
section considered. If the section is chosen to be away from the
corresponding tube-like region, the diagram of the MAE will be
just the same as for the usual solution for a rotating black hole.
If the section goes through the axial singularity, the tube-like
hole in the horizon leaves a trace on all patches of the MAE. The
$r_+$ and $r_-$ surfaces are deformed and approach towards each
other, joining at some distance from the axial singularity and
forming the tube-like channels connecting the interior and the
exterior at some angular direction will appear on all patches of
the diagram.

These {\it black holes with holes in the horizon} have thus
{\it preferred directions} along which the causal structure differs from
that of ``true black holes". Their singularity is, therefore,
naked, but the nakedness is of a very peculiar type, since it
manifests itself in specific directions only. A similar situation
occurs with other non-spherical exact solutions, like e.g. the so
called Gamma metric \cite{hmm}.

\medskip

{\bf 4. Possible astrophysical consequences.}

\medskip

Axial singularities carry travelling electromagnetic and
gravitational waves which propagate along them as along a
waveguide, a phenomenon described by exact singular $pp-$wave
solutions of the Einstein-Maxwell field equations \cite{BurAxi}.
The appearance of the axial singularities in rotating
astrophysical sources may be related to their {\it excitations} by
gravitational and/or electromagnetic waves, and has to be
necessarily caused by some non-stationary process. It was argued
in \cite{BEHM1,BurAxi} that e.m. excitation of black holes leads
inevitably to the appearance of axial singularities.

 The simplest wave modes \be\psi _n = q Y^n
\exp {i\omega _n \t} \equiv q (\tan \frac \theta 2)^n \exp
{i(n\phi + \omega _n \t)} \ee can be labelled by the index $n=\pm
1, \pm 2, ...$, which corresponds to the winding number for the
phase wrapped around the axial singularity. The leading wave terms
have the form $ \cF |_{wave} =f_R \ d \z \wedge d u  + f_L \ d \Z
\wedge d v ,$ where $f_R = (AZ),_1$ and $f_L =2Y \psi (Z/P)^2 +
Y^2 (AZ),_1$ are the factors describing the ``left'' and ``right"
waves propagating, correspondingly, along the $z^-$ and $z^+$
semi-axes.
 Near the $z^+$ axis,  $|Y|\to 0$,
and for $r \to \infty $, we have $Y \simeq  e^{i\phi} \frac \rho
{2r}$, where $\rho$ is the distance from the $z^+$ axis.
Similarly, near the $z^-$ axis $Y \simeq  e^{i\phi} \frac {2r}
\rho  $ and $|Y|\to \infty$. For $|n|>1$ the solutions contain
axial singularities which do not fall off  asymptotically, but are
increasing, denoting instability. For example, the leading wave
for $n=-1$, \be \cF^+_{-1}= - \frac {4q e^{-i2\phi+i\omega _{-1}
\t_+ }} {\rho ^2} \ d \z \wedge du , \ee is singular at the $z^+$
semi-axis and propagates to $z=+\infty.$    The wave excitations
of the Kerr geometry may lead to the appearance of two singular
$pp-$waves which propagate outward along the axial singularities.
In real situations, axial singularities cannot be stable and they
will presumably correspond to some type of jet or burst, and hence
it is natural to conjecture that the related holes in the horizon
may be at the origin of jet formation.

Observational evidence shows a preference for two-jet-like sources,
as e.g., in the field of radio loud sources
\cite{punsly2001,antonnuci1993}.
These jets are emitted in opposite directions along the same axis.
This scenario corresponds to the sum of two singular modes with
$n=\pm 1$
and to two opposite positioned holes in the horizon,  see Fig.3.

\begin{figure}[htb]
\centerline{\epsfig{figure=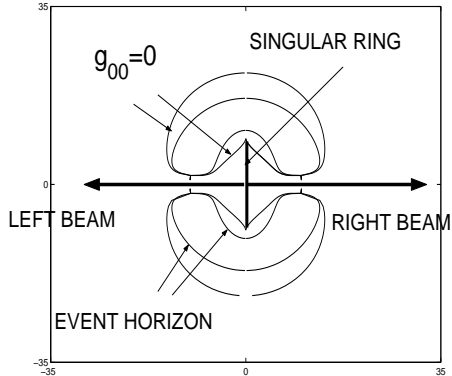,height=5.1cm,width=6cm}}
\caption{{\protect\small  Bi-directional singular beams forming
two holes in the horizon. Parameters: $m=10, \ a=9, \ q=0.3$. }}
\end{figure}

 Electromagnetic $pp-$waves along the singularity
will cause a strong longitudinal pressure pointed outwards from the
hole. It can be easily estimated for the modes of the $pp-$waves with
$n=\pm 1$ \cite{BurAxi}.
For example, the corresponding energy-momentum tensor
is $ T^\mn = \frac 1 {8\pi} |\cF^+_{-1}|^2 k^\m k^n ,$ and the wave
beam with mode $n=-1$, propagating along the $z^+$ half-axis, will
exert
the pressure \be p_{z^+} = \frac {2q^2 e^{ 2 a \omega _{-1}}} {\pi
\rho ^4} , \ee where $\rho$ is an axial distance from the
singularity and $\omega _{-1}$ is the frequency of this mode. For the
exact stationary Kerr-Schild solutions, one can use this expression
in the limit $\omega_{-1}=0$. \medskip

\medskip{\bf 5. Conclusions.}

\medskip

From the analysis above, we conclude that the aligned excitations
of the rotating black hole (or naked rotating source) lead,
unavoidably, to the appearance of axial singularities accompanied
by outgoing travelling waves and also to the formation of holes at
the horizon, which on its turn can lead to the production of
astrophysical jets \cite{jets}.

Multiparticle Kerr-Schild solutions \cite{Multiks}
suggest that axial singularities will
 be bi-directional and oriented along the line connecting
the interacting particles. Thus, it will be interesting to analyze
in further detail the observed jets in order to check the conjecture
that they may be indeed
triggered by radiation coming from  remote active objects.

 As far as the axial singularity survives for
arbitrarily small $q/m \ne0,$ one can consider also the case of
small quantum excitations. One expects that by  elementary quantum
excitations the axial singularities shall lead to the formation of
very small quantum holes with a (maybe just momentary) opening of
the horizon.
This effect may be related to the origin of Hawking's quantum
evaporation.

\medskip

{\bf Acknowledgments.} AB has been supported by the RFBR project
04-0217015-a. EE has been supported by DGICYT (Spain), project
BFM2003-00620 and SEEU grant PR2004-0126, and by AGAUR (Generalitat
de Catalu\-nya), contract 2005SGR-00790. 

\end{document}